\documentclass[pre,twocolumn,showpacs]{revtex4}
\usepackage{graphicx,amsmath,latexsym} 
\begin{document}
\title{Self-consistent mode-coupling theory for the viscosity of
rod-like polyelectrolyte solutions}
\author{Kunimasa Miyazaki}
\email{miyazaki@fas.harvard.edu}
\affiliation{Department of Chemistry and Chemical biology, Harvard
University, 12 Oxford Street, Cambridge MA 02138}
\author{Biman Bagchi}
\affiliation{Solid State and Structural Chemistry Unit, Indian
Institute of Science, Bangalore 560 012, INDIA}
\author{Arun Yethiraj}
\email{yethiraj@chem.wisc.edu}
\affiliation{Theoretical Chemistry Institute and Department of
Chemistry, University of Wisconsin, Madison, WI 53706}

\begin{abstract}
A self-consistent mode-coupling theory is presented for the viscosity
of solutions of charged rod-like polymers. The static structure factor
used in the theory is obtained from polymer integral equation theory;
the Debye-H\"{u}ckel approximation is inadequate even at low
concentrations. The theory predicts a non-monotonic dependence of the
reduced excess viscosity, $\eta_R$, on concentration from the behaviour
of the static structure factor in polyelectrolyte solutions.  The
theory predicts that the peak in $\eta_R$ occurs at concentrations
slightly lower than the overlap threshold concentration, $c^\ast$.  The
peak height increases dramatically with increasing molecular weight and
decreases with increased concentrations of added salt.  The position of
the peak, as a function of concentration divided by $c^\ast$ is
independent of salt concentration or molecular weight. The predictions
can be tested experimentally.
\end {abstract}
\pacs{82.70.Dd;61.25.Hq;61.20.Gy}
\maketitle

\section{Introduction}
\renewcommand{\theequation}{\Roman{section}.\arabic{equation}}  
\setcounter{equation}{0}

Polyelectrolyte solutions are widely considered to be one of the least
understood substances in polymer science \cite{revfs}. There are
several features of these solutions that make them rather complex.  For
one, the long-ranged nature of the electrostatic interactions results
in long-ranged correlations even in dilute solutions.  In addition,
polymer conformations are very sensitive to concentration and ionic
strength because the electrostatic interactions compete with
short-ranged ``hydrophobic" interactions.  This complex static
behaviour is accompanied by very interesting dynamical behaviour of
polyelectrolyte solutions. In this work, we present a theoretical study
of the viscosity of dilute and semi-dilute rod-like polyelectrolyte
solutions using a liquid state approach.

The viscosity of polyelectrolyte solutions displays an interesting
``anomalous" concentration dependence at low polymer concentrations
\cite{vink,cohen}. The quantity that is normally discussed is the
reduced viscosity, $\eta_{R}$, defined as $\eta_{R} \equiv (\eta -
\eta_0)/(\eta_0 c_p)$, where $\eta$ is the viscosity of the solution,
$\eta_0$ is the viscosity of the solvent in the absence of the polymer,
and $c_p$ is the (monomer) concentration of polymers.  Experiments show
that $\eta_{R}$ displays a sharp peak at low polyelectrolyte
concentrations for a variety of different solutions.  This anomalous
concentration dependence of viscosity of dilute polyelectrolyte
solutions has been the focus of attention for over 50 years and,
although there have been many theories that address the problem, it is
not considered to be well understood \cite{revfs,colby1,colby2}.  In
fact, as discussed eloquently by Forster and Schmidt \cite{revfs}, this
problem has witnessed some grave mistakes, causing much confusion.

A contributing factor to the above-mentioned confusion is the fact that
a review of experimental data on the viscosity of polyelectrolyte
solutions shows large inconsistencies \cite{colby1,colby2}. It is only
recently that the reason for this discrepancy between different
experimental measurements has been established.  In careful
measurements of the shear rate dependence of the viscosity, Boris and
Colby \cite{colby1} and Krause et al. \cite{colby2} showed that
polyelectrolyte solutions were shear thinning at extremely low shear
rates, and argued that most of the older experiments did not report the
relevant viscosity in the low shear limit.  

In the last ten years, several theories studies have been put forward.
Notable among them are the mode coupling theory calculation by Borsali
et al. \cite{borsali} and by Rabin et al.\cite{cohen}, the scaling
theory of Dobrynin et al. \cite{andrei}, the effective medium theory of
Muthukumar \cite{muthu}, and the Kirkwood theory of Nishida et al.
\cite{nishida}.  The scaling theory does not predict a peak in the
reduced viscosity, in salt-free solutions.  The theory of Muthukumar
argues that the peak in $\eta_R$ arises from a screening of
intra-molecular hydrodynamic interactions as the concentration is
increased.  The theories of Borsali et al. \cite{borsali}, Rabin et
al.\cite{cohen}, and Nishida et al. \cite{nishida} predict that the
peak in $\eta_R$ arises due to increased screening from counterions as
the concentration is increased, and are similar in spirit to the
mode-coupling theory for charged colloids \cite{hessklein} which is
argued to be accurate for the viscosity of spherical polyelectrolytes
\cite{grohn}. 
Other theories rely on conformational changes of polymers
with changing concentration.  These theories do not take into account
the interesting behaviour in the static structure factor in dilute
polyelectrolyte solutions.

The basic idea of the present work is that the features in $\eta_R$
arise from the behaviour of the static structure factor of dilute and
semi-dilute polyelectrolyte solutions.  In dilute solutions, the static
structure factor displays a prominent peak at low wavevectors
\cite{weber1,weber2}. As the concentration is {\em increased} the peak
{\em broadens} and moves to higher wavevectors (see Fig.\ref{Sk} and
discussion in Section III). This indicates the {\em presence of strong
liquid like correlations on long length scales at low concentrations;
correlations that become less important as the concentration is
increased.}  This observation naturally leads to the question: Could
this strong  non-monotonic concentration dependence of static pair
correlations be the main physics behind the anomalous behaviour of the
viscosity?  To this end, we develop a liquid state theory that
incorporates the behaviour of the static structure factor of {\em
polyelectrolyte} solutions. Such a theory, to our knowledge, has not
been developed previously. 
While it has been suggested 
\cite{antonietti,batzill} that the interesting concentration dependence
of viscosity in dilute polyelectrolyte solutions could arise from the
intermolecular pair correlations (as reflected in the peak in the
structure factor at small wavenumbers), the relationship between the
viscosity and the structure factor is by no means obvious. It is of
interest, therefore to develop a quantitative theory relating the
static structure to the viscosity.

We consider a system of charged rods and present a self-consistent
mode-coupling theory for the viscosity.  We choose to study rods in
order to focus on inter-molecular effects.  There have been many
theories that explain the concentration dependence of the reduced
viscosity on intra-molecular effects.  Studying a system with rigid
molecules allows us to isolate inter-molecular effects since
conformational changes and intra-molecular interactions are absent.
Although we are not aware of experiments for the viscosity of rod-like
polyelectrolytes, these are certainly possible, for example on
solutions of tobacco mosaic virus (TMV) particles.

We present a self-consistent mode-coupling theory for the viscosity of
un-entangled polyelectrolyte solutions.  Starting with the polymer
centre-of-mass density as the slow variable, we develop an expression
for the polymer contribution to the viscosity that is identical to that
of Geszti\cite{geszti}. The intermediate scattering function is
obtained from the self-consistent mode coupling theory as in the
approach of G\"{o}tze and co-workers \cite{franosch1997}. The
centre-of-mass structure factor, required in the theory, is obtained
approximately from integral equation theory\cite{tmvprl,sktmv,ayprl}. 

The theory explains the behaviour of the viscosity based purely on the
behaviour of the static structure factor.  For short chains, the theory
predicts that in salt-free solutions $\eta_{R}$ decreases monotonically
with increasing concentration.  With small amounts  of added salt
$\eta_{R}$ displays a shallow peak as a function of concentration at
low concentrations. For longer chains, the theory predicts a peak in
$\eta_R$ as a function of concentration for all salt concentrations,
including salt-free solutions.  The peak occurs at concentrations
slightly lower than the overlap threshold concentration, $c^\ast$. With
the addition of salt, the intensity of the peak diminishes, but the
position is unchanged.  For a given salt concentration, the height of
the peak increases dramatically as the degree of polymerization is
increased, but the position is unchanged. These predictions can be
tested experimentally.

The rest of this paper is organized as follows:  section II outlines
the theory, section III presents some results and a discussion, and
section IV presents some conclusions.

\section{Mode coupling theory}
\renewcommand{\theequation}{\Roman{section}.\arabic{equation}}  
\setcounter{equation}{0}

The polyelectrolyte solution consists of charged rod-like polymers and
counterions.  Each rod consists of $N$ tangent charged hard spheres of
diameter $\sigma$; counterions are charged hard spheres of diameter
$\sigma$.
We combine a recently developed quantitatively accurate theory for the
liquid structure in charged interacting polyelectrolytes
\cite{tmvprl,sktmv,ayprl} with a self-consistent mode coupling theory
(MCT) to study the dynamics. The starting point for the calculation of
viscosity, $\eta$, is the Green-Kubo formula \cite{balzop}
\begin{equation}
\eta=
\lim_{k \rightarrow 0}
\frac{1}{{k_{\mbox{\scriptsize B}}} TV}\int_{0}^{\infty}\!\!dt~
\langle \sigma^{zx}(k,t) \sigma^{zx}(-k,0) \rangle \ ,
\end{equation}
where ${k_{\mbox{\scriptsize B}}}$ is Boltzmann's constant, $T$ is the
temperature, $V$ is the volume, $\sigma^{zx}(k,t)$ is the transverse
(or off-diagonal) component of the wavevector ($k$) and time ($t$)
dependent stress tensor, and $\langle \cdots \rangle$ denotes an
average over an equilibrium ensemble. The total transverse stress
tensor of a polyelectrolyte solution contains contributions from
solvent, polymer, and small ions. 

For dilute and semi-dilute polyelectrolyte solutions several
simplifications are possible. First of all, the contribution of the
solvent is simply given by the viscosity in the absence of the solute,
$\eta_0$. It is because the presence of low concentrations of polymer
and electrolyte are not expected to alter the solvent dynamics.
Secondly, there is a contribution of the rotational Brownian
motion\cite{doie}.  
Lastly, there is a contribution due to
polymer-polymer interactions $\eta_{p-p}$ which is expected to dominate
over the contributions from small ions. Therefore we argue that $\eta
\approx \eta_0 + \eta_r + \eta_{p-p}$ and focus on the calculation of
the polymer contribution. $\eta_r$ is calculated by neglecting the
effect of the interactions on the rotational degree of freedom and is
given by\cite{doie} 
\begin{equation}
\eta_r = \frac{2}{15}\frac{c_{p}}{N}\zeta_r ,
\end{equation}
where $\zeta_r$ is the rotational friction coefficient and is evaluated
by a hydrodynamic calculation of an ellipsoid of the aspect ratio $N$
as\cite{happel1963} 
\begin{equation}
\zeta_{r}= \frac{2\pi\eta \sigma^3}{3}
\frac{N^4-1}
{\frac{2N^2-1}{\sqrt{N^2-1}}
\ln\left( N + \sqrt{N^2-1}\right)-N}
.
\label{zetar}
\end{equation} 
For $\eta_{p-p}$, we employ a mode coupling treatment similar to that of
Geszti \cite{geszti} to derive a microscopic expression. 

The first step in the mode coupling approach is the choice of the slow
collective variables for the description of the dynamics of the
required correlation functions. Natural choice is the hydrodynamic
variables, {\it i.e.}, the three momentum current densities of polyion,
$J_\alpha({\bf k})$, for the co-ordinates $\alpha$=$x,y,$ and $z$, and
the polyion number density, $\rho_P({\bf k})$, defined as
$\rho_{P}({\bf k})=\sum_{i=1}^{N_{P}} e^{-i{\bf k}\cdot {\bf r}_{i}}$
where ${\bf r}_i$ is the position of the {\em center of mass} of the
$i^{th}$ polyion and $N_P$ is the number of polyions. A calculable
microscopic relation for the viscosity is obtained by using the
projection-operator formalism to re-write the well-known Green-Kubo
time correlation function expression in terms of $P$ and $Q$ operators
\cite{balzop}. The standard approximation in the mode-mode coupling
expansion is to consider the subspace of various binary products of the
basic slow variables. Among such binary products, the odd ones with
respect to time inversion do not contribute to the viscosity, and only
the even combinations can be retained. The two obvious choices of the
binary product are the density-density term and the current-current
term. The current terms are expected to decay much faster than the
density term, due to the friction with the surrounding solvent
molecules. Thus, we neglect this contribution. Finally, all
four-particle correlations are approximated as the product of
two-particle correlations. With the above approximations and
simplifications, the final expression for the zero frequency viscosity
is written in terms of the density correlation function of the polyion
as  
\begin{equation}
\eta = \eta_0 + \eta_r + \frac{{k_{\mbox{\scriptsize B}}} T}{60 \pi^2}
\int_0^{\infty} dk k^4 \int_0^\infty dt 
\left\{
\frac{S^\prime(k)}{S(k)} \right\}^2
\left\{\frac{F(k,t)}{S(k)} \right\}^2,
\label{geszti}
\end{equation}
where $S(k)$ is the static (center-of-mass) structure factor of the
polyions, $F(k,t)$ is the corresponding intermediate scattering
function, $S^{\prime}(k)$ is the derivative of $S(k)$ with respect
to $k$.

In order to evaluate the viscosity, we need the intermediate scattering
function $F(k, t)$ for the polyions and it should also be evaluated
using MCT. As discussed above, we again assume that dynamics of
counterions and solvent molecules is decoupled from dynamics of
polyion. Then, the equation for $F(k, t)$ is expressed in a closed form
as\cite{franosch1997}
\begin{equation}
\frac{\partial ~}{\partial t}
F(k, t)
= 
-\frac{D_{0}k^2}{S(k)}F(k, t) 
- \int_{0}^{t}\!\!{\mbox{d}} t'~ 
  M(k, t-t')\frac{\partial ~}{\partial t'}F(k, t'),
\label{collmct}
\end{equation}
where $D_{0}$ is the bare collective diffusion coefficient. $M(k,t)$ 
is the memory kernel given by 
\begin{equation}
M(k, t) 
=
\frac{ \rho_{P}D_{0}}{2}
\int\!\!\frac{{\mbox{d}}{\bf q}}{(2\pi)^3}~
V_{{\bf k}}^2({\bf q},{\bf k}-{\bf q})
F(|{\bf k}-{\bf q}|, t)F(q, t),
\label{coll.2.Kernel}
\end{equation}
where $\rho_{P} \equiv \langle \rho_{P}({\bf k=0}) \rangle = c_{p}/N$
is the average number density of polyion and $V({\bf q}, {\bf k}-{\bf
q})$ is the vertex function given in terms of the direct correlation
function $c(q)$ as 
\begin{equation} 
V({\bf q}, {\bf k}-{\bf q})
={\hat{\bf k}}\cdot{\bf q} c(q)+{\hat{\bf k}}\cdot({\bf k}-{\bf q})
c(|{\bf k}-{\bf q}|). 
\end{equation} 
Eq.(\ref{collmct}) is a standard MCT equation familiar in the in the
supercooled liquids and
colloids\cite{franosch1997,fuchs1991,banchio2000} community. This
equation is a nonlinear integro-differential equation which has to be
solved self-consistently. The numerical procedure to solve
eq.(\ref{collmct}) following the method proposed in
Ref.[\onlinecite{fuchs1991}].

The bare diffusion coefficient $D_{0}$ is obtained
from the value for a long ellipsoid calculated from 
hydrodynamic calculations using stick boundary conditions.
For the ellipsoid with the aspect ratio of $N$, the total 
diffusion coefficient is given by\cite{happel1963} 
\begin{equation}
D_{0}
= \frac{D_{\parallel}+2D_{\perp}}{3}
= 
\frac{{k_{\mbox{\scriptsize B}}} T}{3\pi\eta_{0}\sigma}
\times
\frac{1}{\sqrt{N^2-1}}
\ln\left(N + \sqrt{N^2-1}\right).
\end{equation}

Note that this theory considers only the translational motion of the
rods, which is assumed to be isotropic.  We argue that we can neglect
the anisotropy in translation and its coupling to rotation in the
concentration regimes we consider. We estimated the contribution from
the anisotropy and coupling to diffusion as follows. In the dilute
limit, if the interaction between polyions is neglected, it is possible
to solve the rotation-coupled diffusion equation and evaluate $F(k, t)$
exactly\cite{miyazakiunpub,maeda1984,kubota1984,doie,footnote}. If the
ratio between the parallel ($D_\parallel$) and perpendicular
($D_{\perp}$) diffusion coefficients is not very large, the change in
the relaxation rate of $F(k,t)$ at short times arising from a coupling
with rotational diffusion is also small. For example, for
$D_{\parallel}/D_{\perp}\simeq 2$, the relaxation rate of $F(k, t)$ at
short times is changed by less than 10 \% due to the coupling with the
rotational diffusion. At longer times, $t \geq \tau_{R}$, where
$\tau_{R}$ is the rotational relaxation time, $F(k, t)$ is simply given
by $\exp[-D_{0}k^2t]$ where $D_0$ is the average diffusion coefficient
defined by $D_{0}= (D_{\parallel}+2D_{\perp})/3$. Thus, decoupling
rotation from translation and assuming the translation diffusion is
isotropic are reasonable approximations in dilute solutions.  These
approximations become questionable when the concentration or rod length
becomes large when, due to the entanglement effects, the rotational
time increases steeply and anisotropy of the translational diffusion
will be enhanced. This regime, however, is far beyond the scope of the
present paper\cite{kuniarun}.

\section{Static properties}
\renewcommand{\theequation}{\Roman{section}.\arabic{equation}}  
\setcounter{equation}{0}

To proceed further we require a model for the polyions and a means of
calculating the static structure of the polyion centers of mass.  In
this work the molecules are modeled as a collection of interaction
sites arranged linearly in a rod-like configuration.  Each particle
consists of $N$ tangent charged hard spheres (or sites) with hard
sphere diameter $\sigma$, which is used as the unit of length in this
paper.  Each sphere carries a negative fractional charge, $f e$, where
$e$ is the charge on an electron.  The effect of solvent and small
ions is included into the potential of interaction between sites on
the polyelectrolyte molecules.  The resulting effective potential,
$\beta u(r)$, is given by 
\begin{equation}
\begin{aligned}
\beta u(r)
= 
\left\{
\begin{array}{c}
\infty \mbox{\hspace{1.25in}for~ $r < \sigma$}
\\
\Gamma \exp(-\kappa r)/r \mbox{\hspace{0.5in}for~ $r > \sigma$}
,
\end{array}
\right.
\end{aligned}
\end{equation}
where $\beta= 1/{k_{\mbox{\scriptsize B}}} T$, $\Gamma = f^2 l_B/(1 +
\kappa \sigma)$, $l_B \equiv \beta e^2/ \varepsilon$ is the Bjerrum
length, $\varepsilon$ is the dielectric constant of the solvent, and
$\kappa$ is the inverse screening length, $\kappa = \sqrt{ 4 \pi l_B
(f^{2} c_p + 2 c_{s}) }$ where $c_s$ is the number density of the
(monovalent) salt, and $c_p$ is the number density of polymer sites. 
In all the calculations presented in this work, $l_B = 0.758 \sigma$,
and $f$=1.  If $\sigma \approx 4$ \AA, then an added salt concentration
of 1 mM corresponds to a reduced salt concentration of 
$c_s \sigma^3 \approx 4 \times 10^{-5}$.
 
The center of mass static structure factor is calculated using integral
equations.  The single chain structure factor, $\hat{\omega}(k)$, is
known exactly for this model.  The site-site static structure factor,
$S_{ss}(k)$, is obtained from the polymer reference interaction site
model (PRISM) theory \cite{prism}, as described elsewhere \cite{sktmv}.
It has previously been established \cite{shew}, by direct comparison of
theoretical predictions for $S_{ss}(k)$ to computer simulations, that
PRISM is accurate for $S_{ss}(k)$. 
The center-of-mass structure factor is the approximated as $S(k)
\approx S_{ss}(k)/\omega(k)$.  To check the validity of this
approximation, we perform Monte Carlo simulations of rods interacting
via screened Coulomb interactions, and calculate $S_{ss}(k)$, and
$S(k)$.  The simulation algorithm is identical to that described
elsewhere \cite{shew} except that we do not perform the Ewald sum.
\begin{figure}
\includegraphics[scale=0.35,angle=-90]{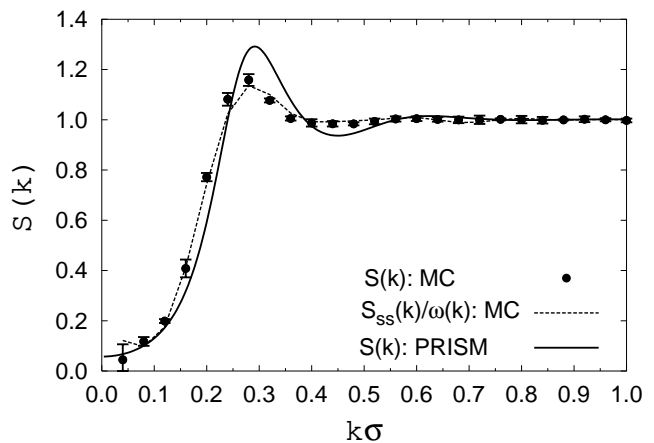}
\caption{Comparison of PRISM predictions (solid lines) for the
centre-of-mass structure factor to Monte Carlo simulation results for
$c_p \sigma^3$=$10^{-3}$, $l_B=\sigma$, and $N$=20. Dashed lines are
simulation results for $S_{ss}(k)/\hat{\omega}(k)$.}\label{mccomp}
\end{figure}
Figure~\ref{mccomp} compares simulations results for $S(k)$ (filled
circles) and $S_{ss}(k)/\omega(k)$ (dotted lines) for $l_B=\sigma$,
$c_p \sigma^3=10^{-3}$, and $N$=20, and shows that the approximation for
$S(k)$ is quite accurate.  Also shown in the figure is the PRISM
prediction for $S(k)$.  The PRISM $S(k)$ correctly reproduces the
liquid like structure manifested in the peak in $S(k)$.  In fact, the
theory is in quantitative agreement with the simulation results. 

\begin{figure}[h]
\includegraphics[scale=0.35,angle=-90]{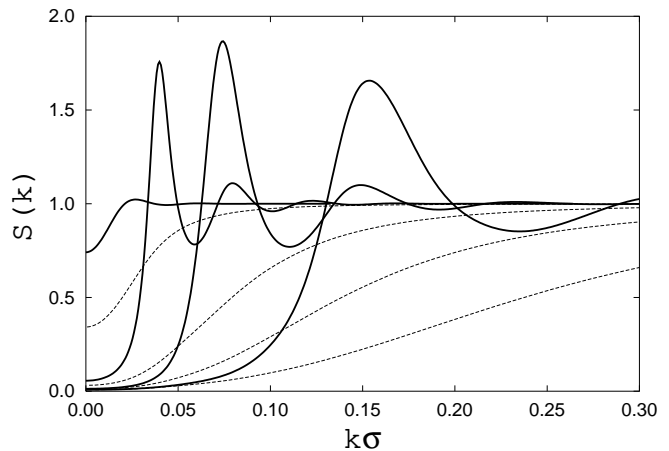}
\caption{Static structure factor, $S(k)$, predicted using the
PRISM theory (solid lines) and the DH approximation (dotted lines) for
$N=150$ and $c_s=1$mM, and for various polyion concentrations. From
left to right, the polyion concentrations are
$c_{p}\sigma^{3}=10^{-6}$, $2\times 10^{-5}$, $10^{-4}$, and $5\times
10^{-4}$.} \label{Sk} 
\end{figure}
Figure \ref{Sk} depicts $S(k)$ from PRISM for $N$=150 and for several
polyion concentrations. We also show the results derived from a simpler
Debye-Huckel (DH) approximation. The DH results is derived by
taking the $\sigma \rightarrow 0$ limit, and approximating the
site-site direct correlation function, $c_{ss}(r)$, by $c_{ss}(r) = -
\beta u(r)$ for all $r$.  The resulting intermolecular structure
factor, denoted $S_{\mbox{\scriptsize DH}}(k)$, is given by 
\begin{equation} 
S_{\mbox{\scriptsize DH}}(k) 
= \cfrac{1}{1+\cfrac{4 \pi l_B}{k^2 + \kappa^2} c_p \omega(k)}.
\label{eq:DHsq} 
\end{equation}
Note that, contrary to the PRISM theory, the DH
approximation is only qualitatively in good agreement at very low
concentrations, where no strong structure is observed.

\section{Results}
\renewcommand{\theequation}{\Roman{section}.\arabic{equation}}  
\setcounter{equation}{0}

For salt-free solutions, the theory predicts that the reduced viscosity
$\eta_R$ is a monotonically decreasing function of polymer
concentration, for short chains.  As the chain length is increased, a
peak in $\eta_R$ is predicted, at concentrations slightly below the
overlap threshold concentration. 
\begin{figure}[h]
\includegraphics[scale=0.35,angle=-90]{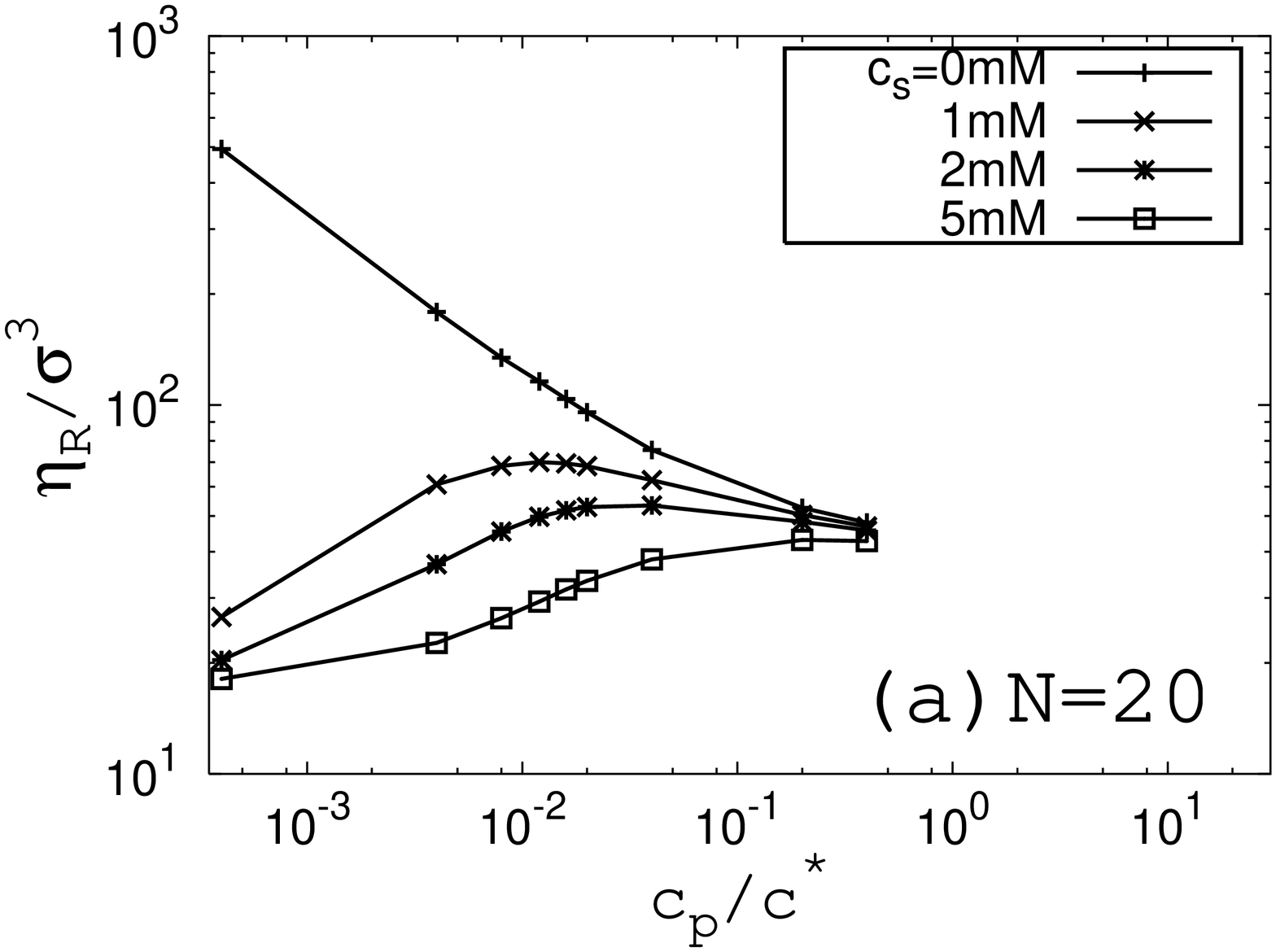}
\includegraphics[scale=0.35,angle=-90]{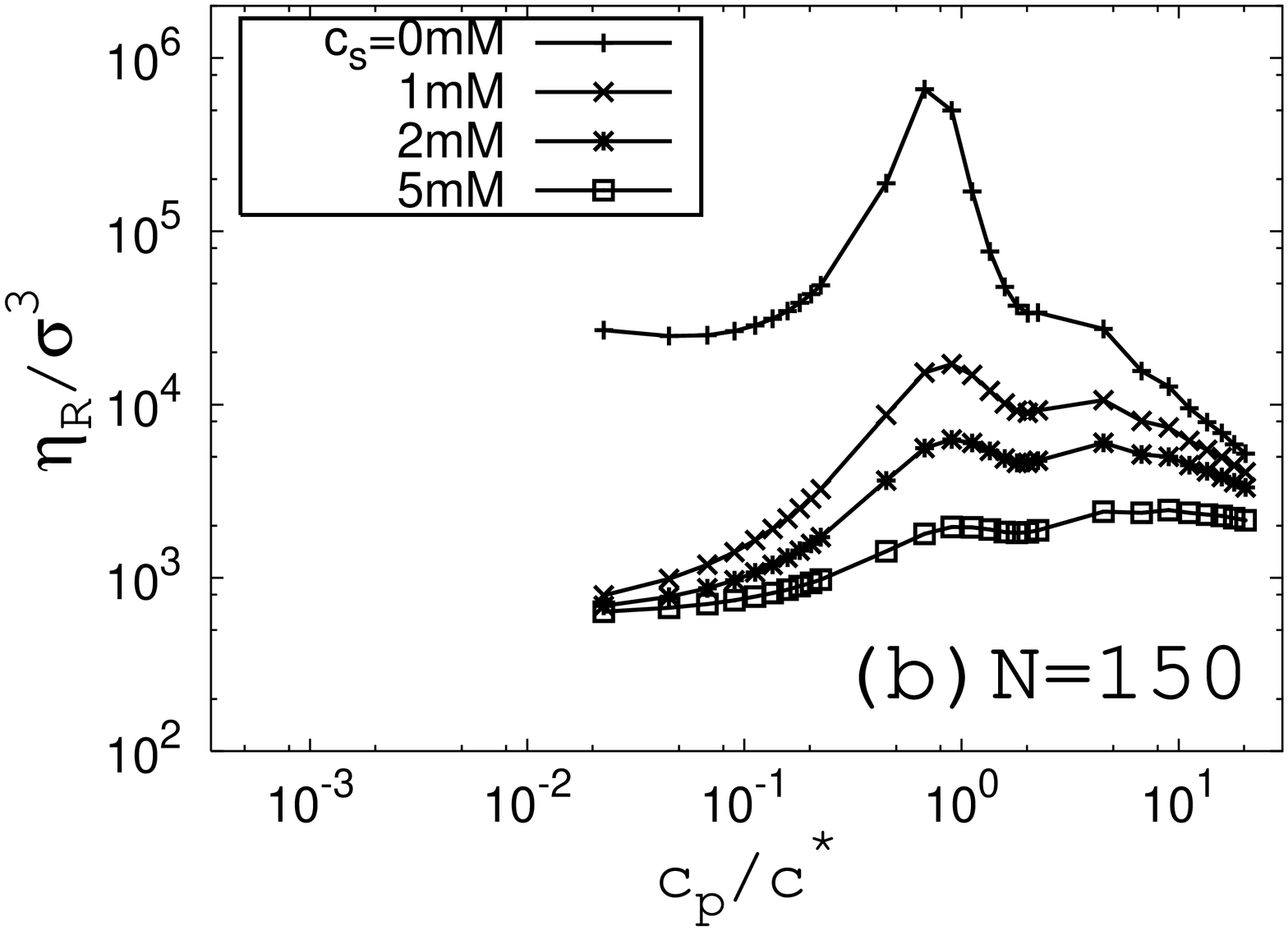}
\caption{Dependence of the reduced viscosity, $\eta_R$, on
polymer concentration for various salt concentrations, $c_s$ = 0 mM
(+), 1 mM ($\times$), 2 mM ($\ast$), and 5 mM ($\Box$), and for (a)
$N$=20 and (b) $N$=150.  Note that the abscissa is concentration
divided by the overlap threshold concentration, $c^\ast$, and both axes
are logarithmic.} \label{saltdep} 
\end{figure}
Figures~\ref{saltdep} (a) and (b)
depict $\eta_R$ as a function of polymer concentration for various
concentrations of added salt, and for degrees of polymerization of
$N$=20 and 150, respectively.  (The added salt concentrations of 1 mM,
2 mM, and 5 mM correspond to reduced salt concentrations of $c_s
\sigma^3$ $\approx$ 4 $\times 10^{-5}$, 8 $\times 10^{-5}$, and 2
$\times 10^{-4}$, respectively.)  In the figures, the abscissa is the
polymer concentration divided by the overlap threshold concentration
$c^\ast$ which, for this model, is given by $c^\ast \sigma^3 = 1/N^2$.
In all cases, we found that the major contribution comes from the
polymer-polymer interaction given by the mode-coupling expression in
eq.(\ref{geszti}). The contribution of the rotational Brownian motion
of the individual rod, $\eta_r$, is independent of the polyion density
and does not affect the qualitative behavior except for the low
concentration regime where the mode-coupling contribution becomes very
small. 

For $N$=20 (Figure \ref{saltdep} (a)) and $c_s = 0$ (salt-free),
$\eta_R$ is a monotonically decreasing function of $c_p/c^\ast$.  As
the salt concentration is increased, the value of $\eta_R$ decreases at
all polymer concentrations. For $N$=20 this results in a shallow peak
in $\eta_R$ at low polymer concentrations and 1 mM salt.  For high
values of added salt $\eta_R$ is a monotonically increasing function of
polymer concentration.  These results are typical of cases when the
static structure factor does not display a very strong peak at low
wave-vectors.  The influence of the long-range liquid like order on the
dynamic properties is therefore very weak.  The predictions for short
chains are qualitatively similar to other theories
\cite{borsali,nishida} that ignore the effect of static structure on
the dynamic properties.

As the degree of polymerization, $N$, is increased, the theory predicts
a prominent peak in $\eta_R$ that occurs at a concentration just below
the overlap threshold concentration.  The amplitude of this peak
increases with increasing degree of polymerization and decreases with
increasing salt concentration.  This can be seen in figure
\ref{saltdep} (b) which depicts $\eta_R$ as a function of $c_p/c^\ast$
for $N$=150.  In salt-free solutions, the peak in $\eta_R$ is very
prominent.  The addition of salt dramatically reduces the height of the
peak, although a peak is clearly present even for high (5 mM) salt
concentrations.  (Note that both axes are plotted a logarithmic scale
in order to fit all the curves on the same figure.)

The reduced viscosity is a strong function of chain length, in a manner
that depends on the salt concentration. 
\begin{figure}[h]
\includegraphics[scale=0.35,angle=-90]{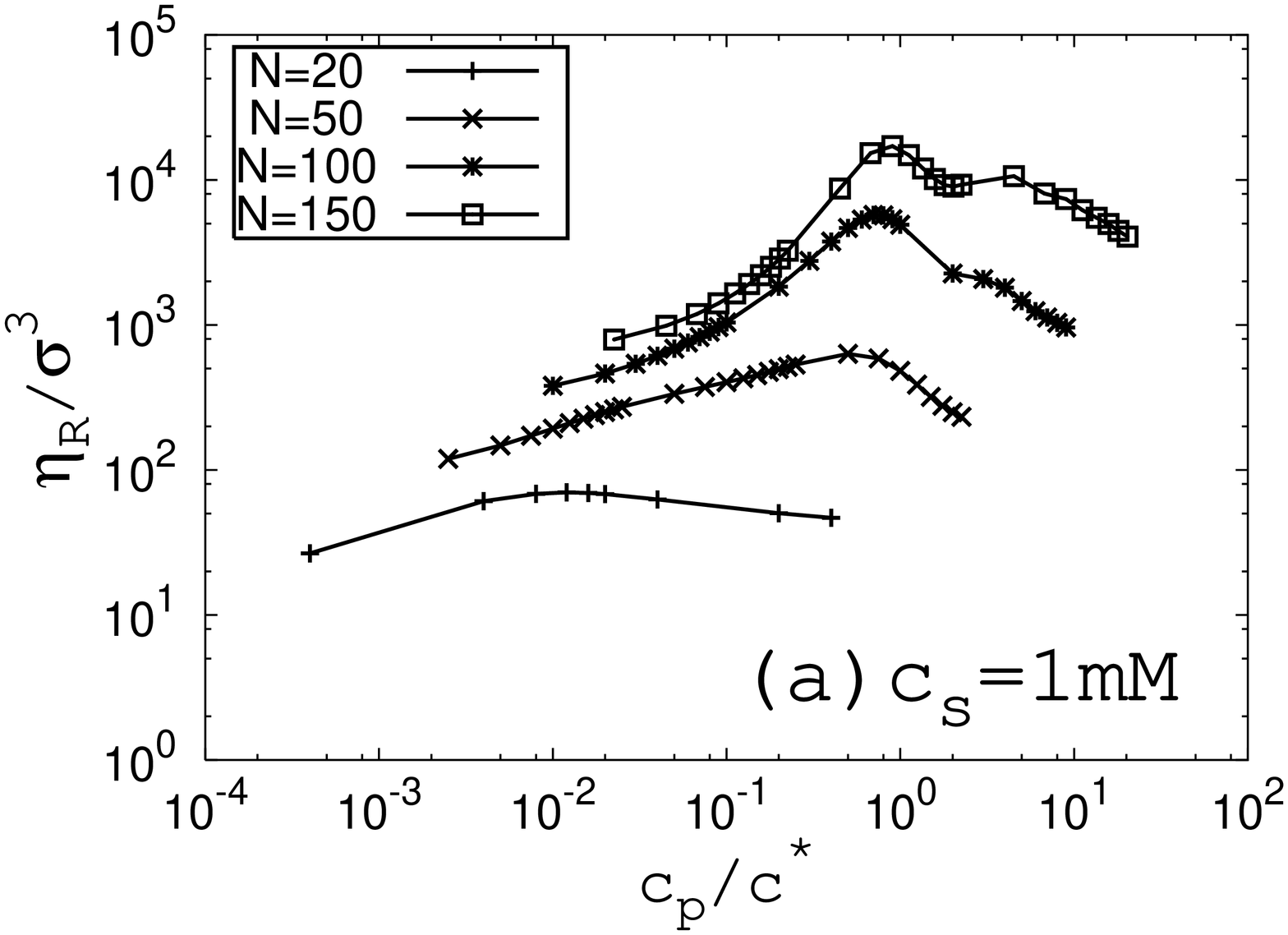}
\includegraphics[scale=0.35,angle=-90]{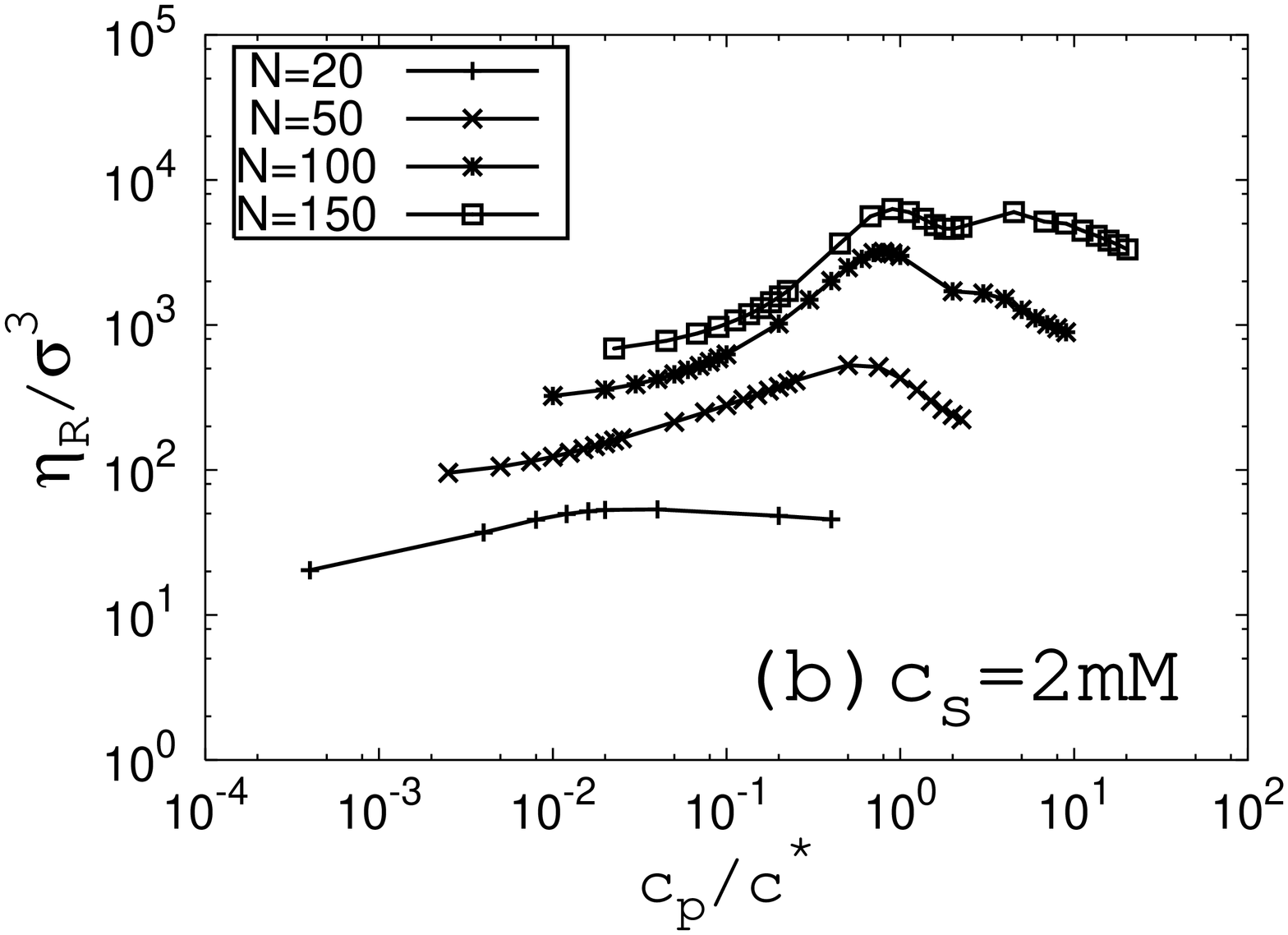}
\caption{Dependence of the reduced viscosity, $\eta_R$, on
polymer concentration for various degrees of polymerization, $N$ = 20
(+), 50 ($\times$), 100 ($\ast$), and 150 ($\Box$), and for (a) $c_s$=1
mM and (b) $c_s$=2 mM.  Note that the abscissa is concentration divided
by the overlap threshold concentration, $c^\ast$, and both axes are
logarithmic.} \label{ndep} 
\end{figure}
Figures \ref{ndep} (a) and (b)
depict $\eta_R$ as a function of polymer concentration for various
values of $N$ and $c_s$ = 1 mM and 2 mM, respectively.  In both
salt-free and added salt solutions, the peak in $\eta_R$ grows with
increasing degree of polymerization, but the position of the peak is
insensitive to the value of $N$.  In salt-free solutions $\eta_R$ is a
strong function of degree of polymerization at low values of
$c_p/c^\ast$ but in added salt solutions $\eta_R$ is almost independent
of degree of polymerization for low values of $c_p/c^\ast$.
\begin{figure}[h]
\includegraphics[scale=0.35,angle=-90]{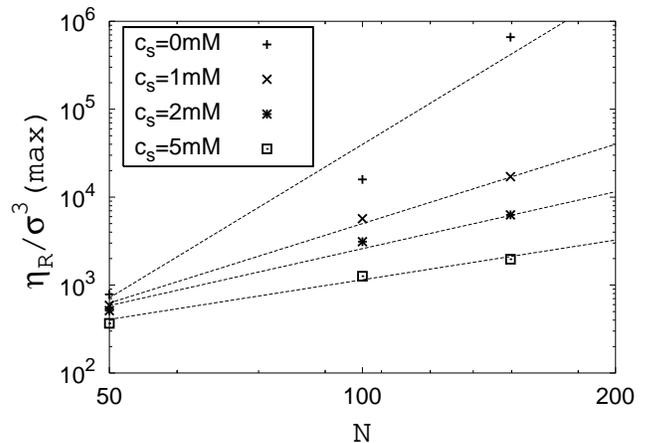}
\caption{$N$-dependence of the peak value of $\eta_{R}$ for
various salt concentrations. Dotted lines are power law fits $N^{\nu}$,
with $\nu$= 5.8, 3, 2.2, and 1.5, for $c_s$ = 0 mM, 1 mM, 2 mM, and 5
mM, respectively.}\label{etarmax}
\end{figure}
Figure~\ref{etarmax} depicts the value of $\eta_R$ at the maximum as a
function of degree of polymerization for various salt concentrations.
The peak values is fitted well by a power law except for the salt-free
case.  The molecular weight dependence is very strong, much stronger
than what is obtained for entangled (neutral) polymer melts.  The
exponent decreases dramatically as the salt concentration is increased.

The physical interpretation of these results is that the peak in
$\eta_R$ arises from intermolecular correlations between the rods.  The
main physical feature that is input into the theory is an accurate
estimate of the static structure factor of the polyelectrolyte
solutions.  The viscosity is then calculated using a fully
self-consistent mode-coupling theory.  Any scaling analysis of the
dependence of $\eta_{R}$ on $N$ and $c_{p}$ must take into account the
complex dependence of static correlations on the dynamics.

\section{Discussion}
\renewcommand{\theequation}{\Roman{section}.\arabic{equation}}  
\setcounter{equation}{0}

The main ingredients of the theory of this work are (i) the use of a
fully self-consistent mode-coupling approximation (SCMCT), and (ii)
accurate estimates of the structure of the solution. It is of
interest to determine how the actual predictions depend on these two
components. We compare the predictions of this work with those of
related, and simpler, theoretical schemes, as described below.  

\noindent
(i) Comparison of SCMCT with lowest order MCT (LMCT).\\
The MCT requires an expression for the intermediate scattering
function, $F(k,t)$, which we obtain from the self-consistent mode
coupling equation, eq.(\ref{collmct}).  A simpler approximation amounts
to neglecting the memory kernel $M(k, t)$ in eq.(\ref{collmct}).  We
refer to this approximation, where $F(k,t)$ is simply given by 
\begin{equation}
F(k,t) = S(k)\exp\left[ - \frac{D_0 k^2 t}{S(k)} \right]
\end{equation}
as the lowest order MCT (LMCT).  Such an approximation has been
previously investigated by others \cite{borsali,cohen} but with the
Debye-H\"{u}ckel (DH) approximation for the static structure. The
theory of Nishida et al. \cite{nishida} is closely related to the LMCT
with an approximate (concentration independent) structure factor
obtained from numerical calculations at zero density. In the LMCT,
eq.(\ref{geszti}) is readily integrated over time to give
\begin{equation}
\eta_{R,\mbox{\scriptsize LMCT}} = 
\frac{k_{\mbox{\scriptsize B}}T }{120 \pi^2 \eta_0 c_p D_0} 
\int_0^\infty\!\!{\mbox{d}}k 
k^2 \frac{ \left[ S^\prime(k) \right]^2 }{S(k)}.
\label{eq:LMCT}
\end{equation}
\begin{figure}[h]
\includegraphics[scale=0.35,angle=-90]{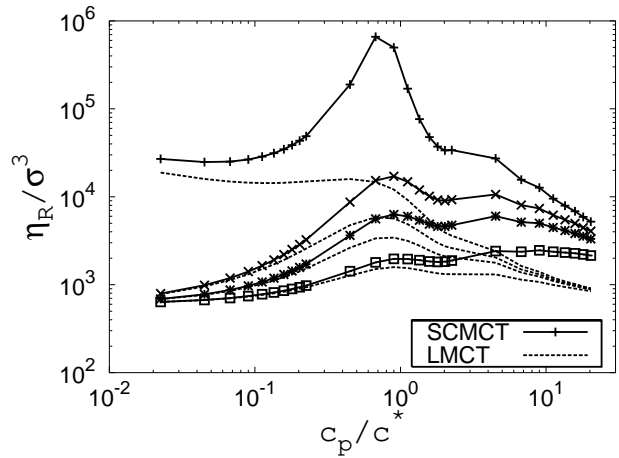}
\caption{Comparison of predictions for the concentration
dependence of $\eta_R$ from a fully self-consistent MCT (SCMCT) (---)
and lowest order MCT (LMCT) (- - -) for $N$=150.  PRISM results for
$S(k)$ are used in all calculations. In each case, the curves
correspond to (from top to bottom) salt concentrations of $c_s$ = 0 mM
(+), 1 mM ($\times$), 2 mM ($\ast$), and 5 mM ($\Box$).} \label{MCTcomp} 
\end{figure}
Figure~\ref{MCTcomp} compares predictions from the SCMCT (solid lines)
and LMCT (dashed lines) for the concentration dependence of $\eta_R$
for $N$=150.  PRISM predictions for $S(k)$ are used in both cases. The
qualitative behaviour predicted by both theories is similar, {\it
i.e.}, there is a peak in $\eta_R$ at concentrations of the order of
the overlap threshold concentration, and the amplitude of the peak
diminishes as the concentration of added salt increases.  The position
of this peak is insensitive to concentration of added salt and degree
of polymerization.  The quantitative differences, however, are
enormous, with SCMCT predicting a value of $\eta_R$ at the peak that is
two orders of magnitude higher than LMCT.  This emphasizes the fact
that memory effects are of considerable importance in these systems.
This difference between LMCT and SCMCT grows with increasing chain
length.  For example the theories are almost indistinguishable for
$N$=20.

\noindent
(ii) Influence of static structure.\\
In order to see how the accurate estimates of the structure affects the
results, we compare MCT predictions for the viscosity using PRISM
results for the structure factor to those using the DH
approximation for $S(k)$ given by eq.(\ref{eq:DHsq}). 
A good test of the importance of liquid structure would be to compare
SCMCT with PRISM $S(k)$ to SCMCT with the DH $S(k)$.  We find, however,
that SCMCT with the DH $S(k)$ predicts a so-called ergodic-nonergodic
transition, i.e., $F(k,t)$ fails to relax to zero, for low polyion
concentrations.  Such a transition, which is also predicted by the MCT
for super-cooled liquids, leads to a divergence in the viscosity at
finite concentrations! This prediction is clearly incorrect, since no
such divergence is seen in experiment, or expected on physical grounds.
This emphasizes, however, that SCMCT is very sensitive to the structure
factor used as input, as one would expect. We attribute the fictitious
transition to an overestimation of the memory kernel of
eq.(\ref{coll.2.Kernel}) at large wavevectors which allows anomalous
positive feedback into the relaxation of $F(k,t)$. This over-estimation
of the memory kernel arises from the broadened and featureless $S(k)$
in the DH approximation.  In reality, the hard-core interaction comes
into play at large $k$, thus resulting in a flattening of $S(k)$.  The
hard-core interaction is, of course, neglected in the DH approximation.

We therefore investigate the influence of structure using the LMCT.
A combination of the DH and LMCT approximations allows us
to derive simple scaling results for the viscosity.  
Since the form factor $\omega(k)$ depends on $k$ only weakly, 
we can set $\omega(k) \approx \omega(0)=N$ in
eq.(\ref{eq:DHsq}) to get 
\begin{equation}
\begin{aligned}
S(k) 
 \approx & \cfrac{1}{1 + \cfrac{4 \pi l_B N c_p \sigma^3}{k^2 +
\kappa^2} } \\
 = & \frac{1}{1 + A \alpha(x)}
,
\end{aligned}
\end{equation}
where $x = k/\kappa$, $A = 4 \pi l_B N c_p \sigma^3/\kappa^2 = N c_p
\sigma^3/(c_p+2 c_s)$,  and $\alpha(x) = 1/(1+x^2)$.  With these
simplifications, $\eta_R$ from the LMCT is given by
\begin{equation}
\eta_{R} =  \frac{ k_B T A^2 \kappa }{120 \pi^2 \eta_0 c_p \sigma^3 D_0}
\int_0^\infty dx x^2 
\frac{ \left[ \alpha^\prime(x) \right]^2 }{(1+A \alpha)^2}
,
\end{equation}
where we have neglected the rotation contribution $\eta_r$, which does
not affect the argument.

If we ignore the concentration dependence of the integral, which is
expected to be weak, then 
\begin{equation}
\eta_R \sim \frac{ N^2 c_p  l_B^2} {D_0 \kappa^3} \sim \frac{N^2
\sqrt{l_B} }{D_0} c_p \left( c_p + 2 c_s \right)^{-3/2}.
\label{eq:etascale}
\end{equation}
For salt-free solutions, this simple scaling approach predicts $\eta_R
\sim 1/\sqrt{c_p}$, which is the same as the Fuoss Law or the scaling
theory of Rubinstein and co-workers \cite{andrei}.  In solutions with
added salt, this scaling approach predicts a peak in $\eta_R$ as a
function of $c_p$.  The peak occurs for $c_p = 4 c_s$, independent of
any of the other parameters.  

\begin{figure}[h]
\includegraphics[scale=0.35,angle=-90]{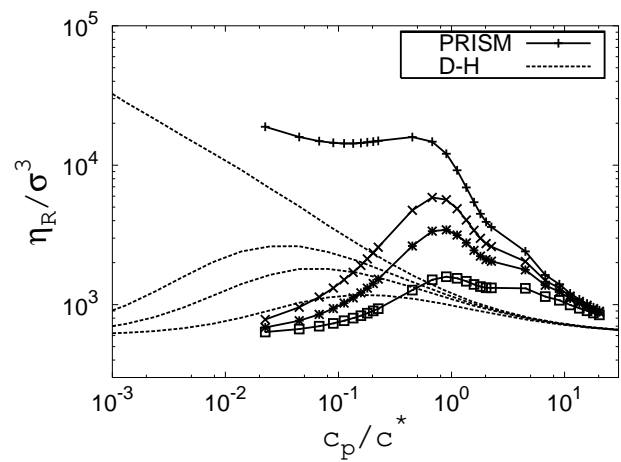}
\caption{Comparison of predictions for the concentration
dependence of $\eta_R$ from lowest order MCT using PRISM input for
$S(k)$ (---) and lowest order MCT using Debye-H\"{u}ckel input for
$S(k)$ (- - -) for $N$=150. In each case, the curves correspond to
(from top to bottom) salt concentrations of $c_s$ = 0 mM (+), 1 mM
($\times$), 2 mM ($\ast$), and 5 mM ($\Box$).} 
\label{sqcomp}
\end{figure}
Figure~\ref{sqcomp} compares predictions for $\eta_R$ using the LMCT,
and DH and PRISM approximations for $S(k)$.  The predictions of LMCT
with DH $S(k)$ is drastically different from that with PRISM $S(k)$. 
For one, no peak in $\eta_R$ is observed in the DH theory in salt-free
solutions.  A peak appears with added salt, but this peak is very broad
compared to the predictions of the theory with PRISM input.  The
position of the peak moves to smaller concentrations as the
concentration of added salt is decreased.  The predictions of LMCT with
DH $S(q)$ is qualitatively similar to the scaling results,
eq.(\ref{eq:etascale}). The physical reason for the peak is completely
different from SCMCT with PRISM and comes from a competition between
the increase in $\eta_R$ with increasing concentration and the decrease
in $\eta_R$ from decreasing the inverse screening length.  The theories
of Cohen et al. \cite{cohen}, Borsali et al. \cite{borsali}, and
Nishida et al. \cite{nishida} are similar in spirit to the lowest order
MCT with DH input.  

\section{Conclusions}
\renewcommand{\theequation}{\Roman{section}.\arabic{equation}}  
\setcounter{equation}{0}

We present a mode-coupling theory for the viscosity of dilute
polyelectrolyte solutions.  Using the static structure factor from
polyelectrolyte integral equation theory as input, we calculate the
viscosity using a self-consistent mode-coupling approach. The theory
predicts a peak in the reduced viscosity for long enough polymer chains
that occurs at concentrations smaller than the overlap threshold
concentration.  When the reduced viscosity is plotted against the
polymer concentration divided by the overlap threshold concentration,
the peak position is independent of degree of polymerization or added
salt concentration, although the height of the peak increases strongly
with increasing degree of polymerization and decreases strongly with
increasing the concentration of added salt.  

An important caveat is that this theory does not attempt to treat
entanglement effects.  The theory assumes that rotational and
translational diffusion are de-coupled, and that the anisotropy in
translational diffusion can be ignored.  These approximations are not
expected to be important in dilute solutions but preclude the
application of this theory to semi-dilute and concentrated solutions of
polyelectrolytes\cite{kuniarun}.

The theoretical predictions are sensitive to the level of approximation
within the MCT and the structure factors used.  If a lowest order MCT
is employed, the theoretical predictions are qualitatively similar to
the self consistent MCT (SCMCT), but the value of the reduced viscosity
is different by an amount that increases dramatically with chain
length.  If the Debye-H\"{u}ckel (DH) approximation is used for the
structure factor, SCMCT predicts a glass transition at low
concentrations, which arises from the neglect of hard-core
interactions.  With the DH structure factor and lowest order MCT
(LMCT), the theory predicts peaks in the reduced viscosity with polymer
concentration, but the position and amplitude of these peaks are
different from what is observed in the full theory.

A rigorous comparison with experiment is not possible because detailed
experiments on the viscosity of rod-like polyelectrolytes in good
solvents are not available.  We therefore compare theory to viscosity
data of (poly)styrene sulfonate (PSS) dissolved in water. Additional
complications in these systems are the possible importance of
intra-molecular effects due to chain flexibility and hydrophobic
interactions.  There are significant differences between theoretical
predictions for rods and experiments on these flexible chains.  In many
experiments \cite{colby2} the peak in $\eta_R$ occurs in the {\em
semi-dilute} regime, while the focus of theoretical work has been on
such a peak in the {\em dilute} solution regime.  
The SCMCT does predict a {\em second} peak in $\eta_R$ 
in the semi-dilute regime (see for example
Figure~\ref{sqcomp}) as well,
but, strictly, this is outside the regime
where we expect the theory to be accurate.
In experiments on sodium PSS
\cite{antonietti}, the peak in $\eta_R$ moves to {\em higher}
concentrations as the molecular weight is increased.  In the SCMCT for
rods the peak position occurs at $c_p = c^\ast$ for all $N$, i.e., the
peak positions moves drastically to lower concentrations as $N$ is
increased.  
We carry out the MCT calculation in the same fashion but
with the center-of-mass structure factor for flexible chains
and find that it gives essentially similar results as rods
and does not explain the $N$-dependence of $\eta_R$.
This is clearly due to the neglect of all other slow modes that
originate from the intra-molecular configuration appearing  in
the case of flexible polymers. 
Muthukumar \cite{muthu} makes a similar prediction. The
colloid-like theories, including those considered by Cohen et al.
\cite{cohen} and Antonietti et al. \cite{antonietti}, predict that the
peak is independent of $N$, and in fact occurs for a concentration $c_p
= 4 c_s$, which is not in agreement with experiment on PSS.  Therefore
none of the theories can claim to explain experimentally observed
results, and the occurrence of the peak in $\eta_R$ must still be
considered ``anomalous".  

In conclusion, we present a fully self-consistent mode-coupling theory
for the viscosity of rod-like polyelectrolyte solutions.  Significant
differences are present between the predictions of the theory and
experiments for {\em flexible} chains. Measurements on theoretically
well understood systems such as solutions of tobacco mosaic virus (TMV)
will therefore be very useful as a test of the theory.  Note that since
the only input into the theory is the static structure factor, which
has been measured in TMV solutions, this theory can be used to make
parameter-free predictions for the absolute viscosity in those systems.
 We hope this work will stimulate further experimental and
computational work on this problem.

\section*{Acknowledgments}

We thank Professor Mathias Fuchs for assistance with the numerical
solution of the MCT equations. K. M. acknowledges support from the
Japan Society for the Promotion of Science (JSPS). A. Y. thanks the
Jawaharlal Nehru Centre for Advanced Research for support and
hospitality during his stay in Bangalore, and Professor Ralph Colby for
useful discussions. This material is based upon work supported by the
National Science Foundation under Grant No. CHE-0315219 (to A. Y.). B.
B. thanks Department of Science and Technology for support.

\end{document}